# Y-Net: A Hybrid Deep Learning Reconstruction Framework for Photoacoustic Imaging *in vivo*

Hengrong Lan[1], *Student Member, IEEE,* Daohuai Jiang[1], Changchun Yang[1], Fei Gao[1,*], *Member, IEEE*

*Abstract*—Photoacoustic imaging (PAI) is an emerging non-invasive imaging modality combining the advantages of deep ultrasound penetration and high optical contrast. Image reconstruction is an essential topic in PAI, which is unfortunately an ill-posed problem due to the complex and unknown optical/acoustic parameters in tissue. Conventional algorithms used in PAI (e.g., delay-and-sum) provide a fast solution while many artifacts remain, especially for linear array probe with limited-view issue. Convolutional neural network (CNN) has shown state-of-the-art results in computer vision, and more and more work based on CNN has been studied in medical image processing recently. In this paper, we present a non-iterative scheme filling the gap between existing direct-processing and post-processing methods, and propose a new framework Y-Net: a CNN architecture to reconstruct the PA image by optimizing both raw data and beamformed images once. The network connected two encoders with one decoder path, which optimally utilizes more information from raw data and beamformed image. The results of the test set showed good performance compared with conventional reconstruction algorithms and other deep learning methods. Our method is also validated with experiments both *in-vitro* and *in vivo*, which still performs better than other existing methods. The proposed Y-Net architecture also has high potential in medical image reconstruction for other imaging modalities beyond PAI.

## I. Introduction

Photoacoustic tomography (PAT) is a kind of hybrid imaging modalities that mixes both optical and ultrasonic imaging advantages. In PAT, ultrasonic wave is excited by a pulsed laser, which has embodied both optical absorption contrast and ultrasonic deep penetration [1-5]. Many practical applications have been investigated to show its great potential in both preclinical and clinical imaging, such as small animal whole body imaging and breast cancer diagnostics [6-15]. Additionally, multispectral PAT has unique advantages in monitoring the functional information of biological tissues, such as blood oxygen saturation ($sO_2$) and metabolism. Specifically, photoacoustic computed tomography (PACT) enables real-time imaging performance, which reveals enormous potential for clinical applications. To obtain the image from the PA signals, image reconstruction algorithm plays an important role. Conventional non-iterative reconstruction algorithms, e.g., filtered back-projection (FBP), delay-and-sum (DAS), are prevalent due to their fast speed. However, the imperfection of conventional algorithms is the existence of artifacts, which results in distorted images, especially in limited view configuration. In this case, the iterative approaches are well adapted with applicable regularization. However, these iteration-based algorithms are time-consuming due to forward operation calculation in every iteration.

In recent years, deep learning has been rapidly developed, especially in computer vision area. Recently, deep learning methods are beginning to attract intensive research interest in image reconstruction problems for medical imaging [16]. The most popular scheme is convolutional neural network (CNN) to post-process the low-quality results from conventional reconstruction [17-24], which has shown satisfactory results. For example, Anas et. al. proposed a new architecture that takes a low quality PA image as input restrains the noise from low power LED-based PA imaging system [25, 26]. Generally, deep learning based non-iterative methods can be divided into two categories: direct processing and post-processing. The difference between them is the format of input data: the former method feeds the raw data and converts into the image at the output of the network; the latter method feeds a poor quality image and converts the feature of the image into the final image. On the other hand, learned iterative schemes train a regularization to optimize the inverse problem [27-30], which need to compute forward and adjoint model alternatively. Unfortunately, its large time consumption still cannot satisfy real-time clinical application, and the number of iterations is restricted by GPUs with limited resources in the training phase.

In this paper, being different from the previous methods, a CNN-based architecture, named Y-Net, is proposed to solve the image reconstruction problem for PACT, which simultaneously has two inputs (measured raw PA signals and rough solution by conventional algorithm) and one output. Particularly, it learned a single reconstruction procedure inspired by iterative schemes (raw PA data and approximate solution as input). This approach fills the gap between existing direct-processing and post-processing methods, which can be called hybrid processing method: both the measured raw data and a beamformed (BF)

Hengrong Lan, Daohuai Jiang, Changchun Yang and Fei Gao are with the Hybrid Imaging System Laboratory, School of Information Science and Technology, ShanghaiTech University, Shanghai 201210, China (*corresponding author: gaofei@shanghaitech.edu.cn).

Hengrong Lan is also with Chinese Academy of Sciences, Shanghai Institute of Microsystem and Information Technology, Shanghai 200050, China, and University of Chinese Academy of Sciences, Beijing 100049, China.

image are used as inputs. These two inputs contain different types of information respectively: rich details and overall textures. In this work, the measured PA signals are acquired by linear array probe, which suffers limited-view problem.

The overview of this paper is arranged as follows. Firstly, we review the physical model of PAT and inverse problem. Then, we generalize the deep learning method to reconstruct the PA image. In Method section, we show a detailed description of the architecture and implementation of our proposed method. In the experiment section, we illustrate the generation of training data and the experimental setup. In Results section, we show the simulation, *in-vitro* and *in-vivo* results compared with conventional reconstruction algorithms and other deep-learning based methods, such as U-Net. Finally, we discuss some details and conclude this work followed by future work. The preliminary results will be present in EMBC 2019 [31].

## II. BACKGROUND

### A. Photoacoustic Imaging

PA wave is excited by a short pulse laser, and we can derive the forward solution based on Green's function. From the PA generation equation, the propagating PA signal in both time and spatial domain $p(\mathbf{r}, t)$ triggered by the initial pressure $p_0(\mathbf{r})$ satisfies [4]:

$$\left(\nabla^2 - \frac{1}{v_s^2}\frac{\partial^2}{\partial t^2}\right)p(\mathbf{r},t) = -p_0(\mathbf{r})\frac{d\delta(t)}{dt}, \quad (1)$$

where $v_s$ is the speed of sound. We can write the forward solution of PA pressure detected by transducer at position $\mathbf{r}_0$ [32]:

$$p_d(\mathbf{r}_0, t) = \frac{\partial}{\partial t}\left[\frac{t}{4\pi}\iint_{|\mathbf{r}_0-\mathbf{r}|=ct} p_0(\mathbf{r})d\Omega\right], \quad (2)$$

where $d\Omega$ is the solid angle of the transducer with respect to the point at $\mathbf{r}_0$. For the PAT inverse problem, the main idea is to reconstruct the initial pressure $p_0(\mathbf{r})$ from the raw PA signals received by transducer $p_d(\mathbf{r}_0, t)$.

The conventional back-projection calculates the inverse equation, which can be expressed as [33]:

$$p_0(\mathbf{r}) = \frac{1}{\Omega_0}\int_{S_0}\left[2p(\mathbf{r}_0,t) - \frac{2t\partial p(\mathbf{r}_0,t)}{\partial t}\right]\frac{\cos\theta_0}{|\mathbf{r}-\mathbf{r}_0|^2}dS_0, \quad (3)$$

where $\theta_0$ is the angle between the vector pointing to the reconstruction point $\mathbf{r}$ and transducer surface.

Let $f = p_0(\mathbf{r})$ and the measured data by sensor equal to $b$, and we use a linear operator $A$ represent the forward model, then we have:

$$Af = b. \quad (4)$$

To solve the inverse problem, the main idea is recovering $f$ from the known $b$.

### B. PA Image Reconstruction

PA image can be reconstructed from the intact raw data by solving Eq. (1). Many pre-clinical applications require real-time imaging performance, which put computation efficiency as a basic requirement for algorithm design. By proper approximation of these wave equations, many beamforming algorithms such as time-domain delay-and-sum (DAS) and time reversal [34-37], have been widely applied in real application due to their fast speed and easy implementation.

In this work, we choose the DAS algorithm to generate the rough PA images as one of the network inputs, which is an approximate solution. Whilst these time-efficient approaches suffer from severe artifacts. Fig. 1 indicates the difference between the images reconstructed by conventional reconstruction and ground-truth, which shows that the DAS reconstructed image loses some backbone information with severe artifacts. Fig. 1(c) is the differential image of Fig. 1(a) and (b) highlighting the major different vessels, most of which are vessels perpendicular to the linear ultrasound array.

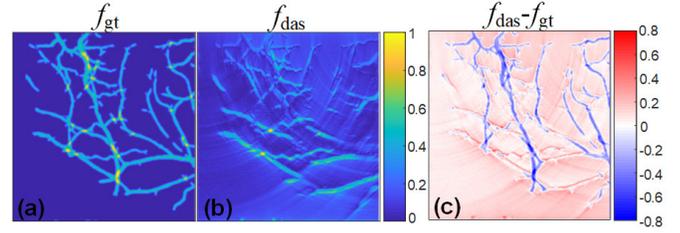

Fig. 1. Comparison of information loss in the traditional DAS reconstruction method. (a) The ground-truth; (b) The delay-and-sum reconstructed result of (a); (c) The difference between (a) and (b).

Model-based approach can reconstruct the imperfect data well compared with above non-iterative algorithms, which devotes to rebuild PA image $f$ from signal $b$ by optimizing the objective function:

$$\arg\min_{f} \frac{1}{2}\|Af - b\|_2^2 + \lambda\mathcal{R}(f), \quad (5)$$

where $\frac{1}{2}\|Af - b\|_2^2$ indicates the data consistency, and the $\mathcal{R}(f)$ is the regularizing term, $\lambda$ is a regularization parameter. It can be solved in many methods iteratively [38-43], which however sacrifices the computation efficiency.

### C. Deep Learning for Reconstruction

Deep-learning-based approach has been developed to resolve the image reconstruction problem. Non-iterative deep-learning-based approaches can be divided into direct and post-processing schemes. The former scheme maps the sensor data $b$ to initial pressure $f$ using a CNN framework, which can be generally expressed as:

$$\arg\min_{\Theta} \mathbb{E}_{b,f}\|\mathcal{N}(\Theta, b) - f\|_2^2. \quad (6)$$

This problem is approximately solved over a training dataset $\{(b_i, f_i)\}_{i=1}^{N}$. However, this method does not contain physical models, and is only driven by data, leading to lower generalization and robustness. On the other hand, the latter scheme considers the approximate solution of physical model and the parameters of network subject to learning are:

$$\arg\min_{\Theta} \mathbb{E}_{f^*,f}\|\mathcal{N}(\Theta, f^*) - f\|_2^2, \quad (7)$$

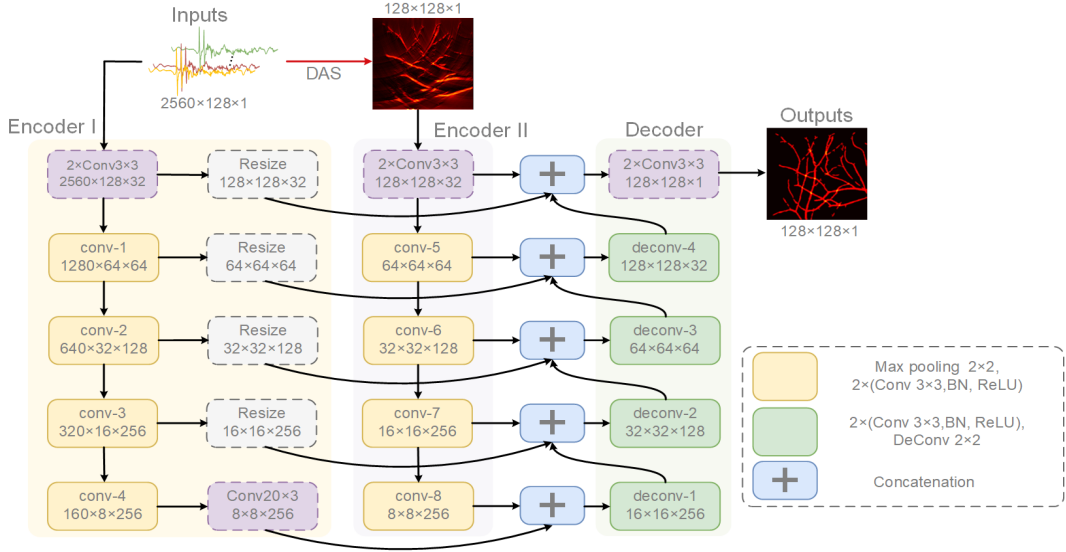

Fig. 2. The architecture of Y-Net. Two encoders extract different input feature, which concatenates into the decoder. Both encoders have skip connections with the decoder. DAS: delay-and-sum; (H×W×C) in blocks specify the output dimension of each component; ConvH×W indicates the convolution operations with H×W kernel size. All operations accompanied by a Batch-Normalization(BN) and a ReLU.

where $f^*$ is the approximate solution generated by conventional non-iterative algorithm, such as DAS. This scheme has rough texture information of the object and shows better performance compared with the previous scheme. However, the detailed information of object may be lost as the input DAS-generated images are imperfect and suffers severe artifacts.

Both abovementioned non-iterative schemes have their own drawback respectively, and current research work mostly focused on ameliorating the neural network. In this paper, we fill the gap between existing two approaches, and propose a new scheme, which fuses and complements each other of the two schemes. To implement this scheme, we propose a new representational framework, named Y-Net, which will be introduced in the next section.

## III. Proposed Method

Most CNN architecture only establishes a single input-output stream for imaging reconstruction (e.g. signals only or image only). Based on above analysis, the scheme with signals' input only or with images' input only suffers their own drawbacks, respectively. Therefore, we assume that it may be a good solution to combine the raw PA signals and beamformed images as input data. It deserves noting that the raw PA signals and beamformed image have different size and features, which inspired us to build the neural network with two inputs.

Our proposed scheme can be termed as hybrid processing, and a pair of inputs are fed into the network to learn the parameters subject to:

$$\arg\min_{\Theta} \mathbb{E}_{(f^*,b),f} \left\| \mathcal{N}(\Theta,b,f^*) - f \right\|_2^2. \quad (8)$$

This scheme incorporates more texture information compared with the direct-processing scheme, and more physical information compared with the post-processing scheme. Since these schemes do not rely on forward models, the proposed method has the ability to satisfy real-time imaging requirements.

The proposed Y-Net integrates both features with two inputs by two different encoders. The global architecture of Y-Net is shown in Fig. 2, which inputs the raw PA signals to an encoder, and processes the raw data to obtain an imperfect beamformed image as the input of another encoder. Being different from U-Net [44], the proposed Y-Net enables two inputs for different types of training data that is optimized for hybrid image reconstruction. The Y-Net consists of two contracting paths and a symmetric expanding path. Encoder I and Encoder II encode the physical features and texture features respectively, and the final decoder concatenates the features of both encoder outputs and generates the final result.

### A. Encoder for measured data

The Encoder for measured data (Encoder I) takes the raw PA signals as input. It is similar to the contracting path of U-Net. An extra 20×3 convolution is put on the middle of the bottom layer, which translates the 160×8 features map to 8×8. Every layer also shared their information with the Decoder mirrored layers by resizing and skipping connection. The raw data contains a complicated feature, and Encoder I filtrates the feature as a supplement for the information loss of reconstructed image during the beamforming process.

The Encoder I maps a given PA signal $b \in \mathbb{R}^{N_b}$ to a features space $z \in \mathbb{R}^{N_k}$. Assuming it only has one convolution every layer of the encoder, we can denote the $i$-th channel of $k$-th layer for Encoder I:

$$\varphi_i^k = \sigma_2\left(\mathrm{P}^{k\mathrm{T}} \sigma_1\left(\sum_{j=1}^{s-1}\left(\varphi_j^{k-1} * \kappa_{i,j}^k\right)\right)\right), i < s. \quad (9)$$

Where $s$ is the output channels size, $\kappa$ is the convolutional kernel, and $\sigma(\cdot)$ is the batch normalization (BN) and rectified linear unit (ReLU) operation, P is pooling operation, * denotes the convolution operation. Furthermore, we also rewrite the matrix representation of the $k$-th layer for double convolution operation:

$$\varphi^k = \sigma_3\left(\sigma_2\left(\sigma_1\left(P^{kT}(\varphi^{k-1})\right) * \kappa_1^{kT}\right) * \kappa_2^{kT}\right), \quad (10)$$

all the operations are matrix operations. For the first layer, the input is measured data, without the pooling operation. We can rewrite the parameterization of Encoder I:

$$z_1 = \left(\varphi^5 \cdots \varphi^1(b)\right) * \kappa^{5T} = E_1(\mathcal{W}_{E1}, b) * \kappa^{5T}, \quad (11)$$

where $\mathcal{W}_{E1}$ is parameter matrices: $\mathcal{W}_{E1} \in \mathbb{R}^{N_5} \times \cdots \times \mathbb{R}^{N_1}$. The kernel $\kappa^5$ with 20×3 size map the feature from 160×8 to 8×8. We do not explicitly tune the bias term since it can be incorporated into $\varphi$. Meanwhile, the signals have a longer size in time-dimension, and a larger receptive field is desirable to focus more information in this dimension. Although $z_1$ is latent features of PA image, most dimensions are asymmetric before last convolution operation. These parameters should be estimated during the training phase.

### B. Encoder for reconstructed image

The Encoder for reconstructed image (Encoder II) takes the image reconstructed from raw PA data by a conventional algorithm (we used the delay-and-sum). The structure of every layer is the same as Encoder I except the bottom layer. Every layer unit is composed of two 3×3 convolutions, BN and ReLU, and a maximizing pooling to downsample the features. The image is passed through a series of layers that gradually downsample, and every layer acquires different information respectively. Meanwhile, every layer shared their information with the decoder mirrored layers by skip connection. It is desirable to concatenate many low-level information such that the location of texture will be passed to the decoder.

Similarly, the Encoder II maps a reconstructed PA image $f^* \in \mathbb{C}^{N^*}$ to a features space $z \in \mathbb{R}^{M_k}$. The matrix representation of the $k$-th layer for Encoder II is similar to Encoder I:

$$\varphi^k = \sigma_3\left(\sigma_2\left(\sigma_1\left(P^{kT}(\varphi^{k-1})\right) * \kappa_1^{kT}\right) * \kappa_2^{kT}\right). \quad (12)$$

For the first layer, the input is reconstructed image without the pooling operation. We can also rewrite the parameterization of Encoder II as:

$$z_2 = \varphi^5 \cdots \varphi^1(f^*) = E_2(\mathcal{W}_{E2}, f^*), \quad (13)$$

where $\mathcal{W}_{E2}$ is parameter matrices: $\mathcal{W}_{E2} \in \mathbb{R}^{M_5} \times \cdots \times \mathbb{R}^{M_1}$. From the E2, the reconstructed image will be encoded as latent features.

### C. Decoder of Y-Net

The outputs of the two encoders are taken to the decoder after concatenation, which is symmetric with Encoder II. Every layer unit is composed of two 3×3 convolutions, and an up-convolution to upsample the features. On the other hand, every layer receives low-level information from two encoders' mirrored layers and concatenate with the feature from previous layer of the decoder. The final layer will generate a 128×128 image.

The decoder takes two feature maps from different encoder as an input, process it and produce an output $f \in \mathbb{C}^N$. For the decoder, every layer is fed by two skipped connections from two encoders except the feature from the prior layer. The corresponding operation at the $k$-th layer encoder is described by:

$$\chi^k = \varphi^k = \sigma_3\left(\sigma_2\left(\sigma_1\left(P^{kT}(\varphi^{k-1})\right) * \kappa_1^{kT}\right) * \kappa_2^{kT}\right), \quad (14)$$

where $\chi^k$ denotes the skipped feature. Particularly, the skipped feature of Encoder I needs to resize to the same dimension with other feature. Similarly, the Decoder maps these features to a final PA image $f \in \mathbb{C}^N$. We also rewrite the matrix representation of the $k$-th layer with skipped connection:

$$\varphi^k = \sigma_3\left(U^{kT}\left(\sigma_2\left(\sigma_1\left(\varphi^{k-1} + R(\chi_1^k) + \chi_2^k\right) * \kappa_1^{kT}\right) * \kappa_2^{kT}\right)\right), \quad (15)$$

where $U(\cdot)$ is up-convolution operation, $R(\cdot)$ is the resizing operation. It is noteworthy that every channel of Decoder layer has triple inputs including two encoder features and prior feature. For the final layer, the output is the final image, without the up-sampling operation. Meanwhile, we can rewrite the parameterization of Decoder as:

$$f = \varphi^5 \cdots \varphi^1(z_1, z_2) = D(\mathcal{W}_D, z_1, z_2), \quad (16)$$

where $\mathcal{W}_D$ is parameter matrices: $\mathcal{W}_D \in \mathbb{C}^{N_5} \times \cdots \times \mathbb{R}^{N_1}$. Two inputs ($z_1$ and $z_2$) are different dimensional features, which are mapped to the final image by $D(\cdot)$.

### D. Implementation

As shown in Fig. 2, every module of convolutions contains BN and ReLU ($f(x) = \max(0, x)$). Encoders and decoder have five layers respectively, and the output size of every layer has been annotated in the block in Fig. 2.

We use the mean squared error (MSE) loss function to evaluate the reconstructed error. Adam optimization algorithm [45] is used to optimize the network iteratively. The MSE loss is defined as:

$$L_{rec}(f) = \frac{1}{2}\|f - gt\|_F^2, \quad (17)$$

where $f$ is the reconstruction image, and $gt$ is the ground truth. In our method, the Encoder II provides the main texture, so we should further penalize Encoder II by an auxiliary loss:

$$L_{aux}(z_2) = \frac{1}{2}\|z_2 * \kappa^T - R(gt)\|_F^2, \quad (18)$$

where $R(\cdot)$ is resizing operation, the channels of $z_2$ convert to one channel by convolution. Finally, we train the network by minimizing the total loss:

$$L_{total} = L_{rec} + \lambda L_{aux}, \quad (19)$$

where $\lambda$ is hyper-parameter, and we chose $\lambda = 0.5$ in the training phase.

Pytorch [46] is used to implement the proposed Y-Net. The hardware platform we used is a high-speed graphics computing workstation consisting of two Intel Xeon E5-2690 (2.6GHz) CPUs and four NVIDIA GTX 1080Ti graphics cards. The batch size is set as 64, and the running time is 0.453 second per batch. The iteration is set as 1000 epochs, and the initial learning rate is 0.005. The source code is available at https://github.com/chenyilan/Y-Net.

## IV. Experiment

### A. Numerical vessels data generation

The deep-learning-based approach is a data-driven method that requires a number of data for training to get the desired results. Unfortunately, PAT does not have access to a large amount of clinical data to train the network as a kind of newly developed imaging technology. Especially for reconstruction problems, we often need raw data, which is usually only available in research lab. Therefore, we seek to train neural networks using simulation data and test the trained models in experiments both *in vitro* and *in vivo*.

The MATLAB toolbox k-Wave [47] is used to generate the training data. The simulation setup is shown in Fig. 3, where a linear array transducer was placed at the top of the region of interest (ROI). The sample is placed in the 38.4×38.4 mm size of ROI, where the linear array probe with 128 elements can receive the PA signals from the sample. We record the raw data from the sensor, generate beamformed images and ground-truth for training and testing. All images have 128×128 pixels, and acoustic speed is set as 1500 m/s. The center frequency of the transducer is set as 7 MHz with 80% fractional bandwidth. We final allot a 2560×128 input size for PA signal, which has 60 dB SNR.

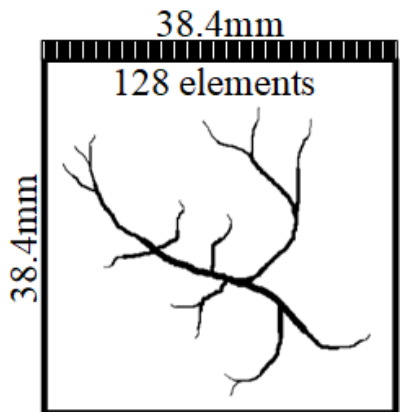

Fig. 3. The illustration of the simulation setup.

The factitious segmented vessel from public fundus oculi DRIVE [48] can be deployed with initial pressure distribution. The vessels need to be segmented and pre-processed: 1). the complete blood vessel of fundus oculi is segmented into four equal parts; 2). randomly rotational transform (90°, 180°, 270°) and superpose two segmented blood vessels. After a series of operations, the excessive dataset will be loaded into k-Wave simulation toolbox as the initial pressure distribution.

The dataset consists of 4700 training sets and 400 test sets, which are generated by MATLAB k-Wave toolbox for PA simulation.

### B. Verification of simulation data

We trained all models on the numerical training data, and verify on the test set. In this phase, we compare our method with ablation study and some existing models as following:
- ➢ Two variant Y-Net, which removes the connection of raw data (Encoder I) and the connection of the beamformed image (Encoder II) with the Decoder respectively.
- ➢ The post-processing method: U-Net [44], the input is the result of DAS image.

We compare our method with the non-iterative learned method in our paper. All learned methods use the same data set and test on other data.

### C. Application to in-vitro data

In order to further verify the feasibility of our proposed method, an *in vitro* phantom was prepared by a chicken breast tissue inserted with two pencil leads. The PACT system is depicted in Fig. 4: a pulsed laser (532 nm, 450 mJ, 10 Hz) illuminates the sample through an optical fiber, and a data acquisition card (DAQ-128, PhotoSound) received and amplified the PA signals from the 128 channels' ultrasound probe (7 MHz, Doppler Inc.). The data sampling rate is 40 MHz, and data length is 2560 points. The system is synchronously controlled by a computer, including laser firing and data acquisition. Two leads are inserted in the chicken breast tissue as "V" shape in Fig. 4, and the ROI is same as the simulation setup.

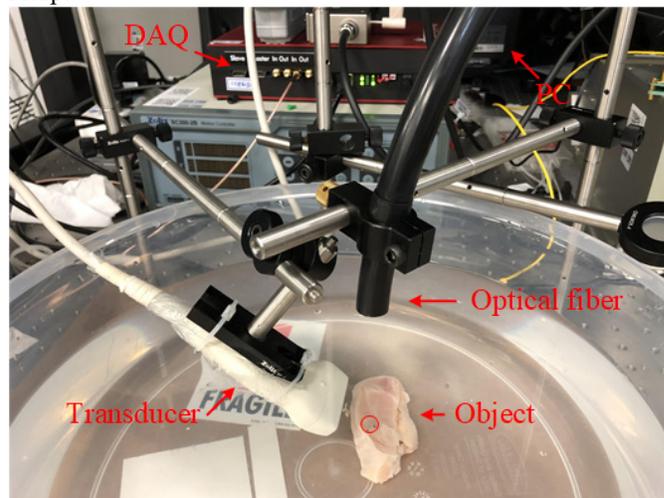

Fig. 4. The schematic of PACT system setup; red circle indicates the pencil lead. DAQ: data acquisition card; PC: personal computer.

### D. Application to human in-vivo data

Last but not least, the *in vivo* PA imaging experiments of a human palm have also been performed to validate our approach. The system setup is the same as the *in-vitro* experiment. Both phantom and *in-vivo* data have different characteristics that are not perfectly represented by the training on synthetic data.

In order to improve the results, we alter the input of Encoder II, which is revised as a better texture reconstructed result instead of DAS. Likewise, the results of post-processing method are also improved. For our architecture, Encoder II provides a main texture of output, and we will obtain an improved texture if the Encoder II is fed with a preferable input images. The Encoder I will supplement the missing information in beforehand reconstruction.

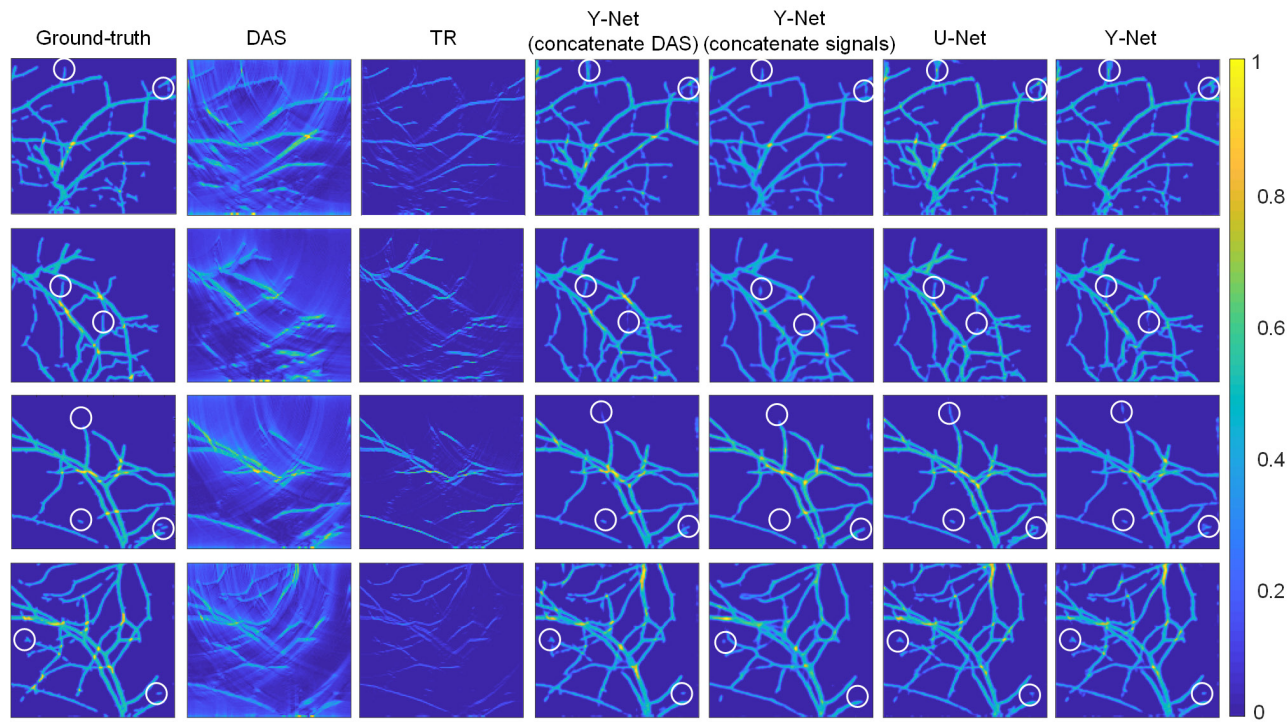

Fig. 5. The example of performance comparison using different methods to reconstruct initial pressure. The four examples correspond to four rows; every column corresponds to different method, from left to right: ground-truth, DAS, TR, Y-Net only concatenate Encoder II into the Decoder, Y-Net only concatenate Encoder I into the Decoder, post-processing U-Net and Y-Net. DAS: delay-and-sum; TR: time reversal. The white circles indicate the local details.

## V. Results

### A. Evaluation of synthetic data

We compared two different conventional algorithms and three different models with our proposed approach. Time reversal (TR) and delay-and-sum (DAS) are selected as conventional algorithms for evaluating performance. To visually compare the performance of different methods, four examples of imaging results from the test set are shown in Fig. 5. From left to right, the method is DAS, TR, Y-Net only concatenating Encoder II into the Decoder, Y-Net only concatenating Encoder I into the Decoder, post-processing U-Net and the proposed complete Y-Net.

The conventional algorithms are easily fooled by artifacts, and we can still see the appearance of the object roughly. The deep-leaning-based approach almost restores the rough outline of the sample, and its performance differs for reconstructing the details of small vessels. From the local details of Fig.5 (white circles), we can see that all models connected to beamformed (BF) input are susceptible to strong artifacts in BF, and introduced some errors in the details. Y-Net (concatenate signals) can avoid the abovementioned errors, but it is difficult to identify the small independent source. The proposed complete Y-Net provides a clearer texture in detail than the U-Net, which indicates that Y-Net is more anti-disturbing to artifacts in BF by integrating the information in raw data. So the performance of Y-Net may be further improved by utilizing more advanced BF algorithm, which seamlessly bridges the joint improvements of conventional reconstruction algorithms and deep learning.

Furthermore, we can analyze the resolution using the point-target, which will help on evaluating these methods from another perspective. Nine points phantom has been placed in two rows as the Fig. 6(a) showed. We compare our method with DAS, TR and U-Net in Fig. 6(b)-(e). In conventional algorithms, many artifacts adjoin the target points in Fig. 6 (b) and (c). In practice, most non-iterative algorithms are unable to eliminate artifacts especially in limited-view configuration. Our method eliminates most artifacts compared with U-Net, but deep-learning-based method can introduce a slight distortion due to the gap between training data and point-like data. Taking a look at a horizontal cross section of the white dotted line, the profile along the white dotted line also indicates the superiority of our method compared with others in Fig.6 (f).

Three indexes for quantitative evaluation are used as the metric to evaluate the performance of different methods:

1). Structural Similarity Index (SSIM) [49], a higher value indicates a better quality for estimated image, which is simply defined as:

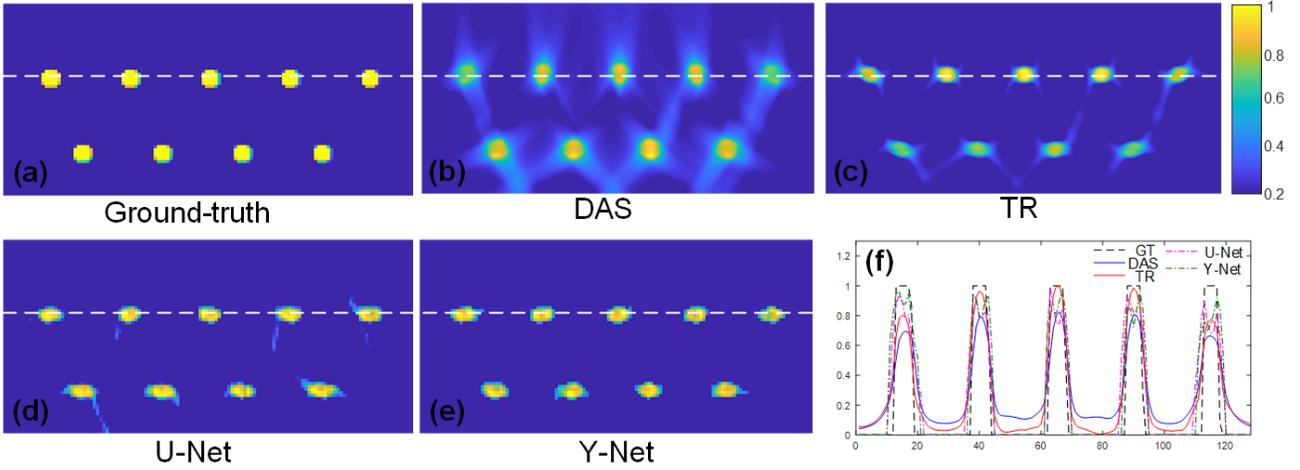

Fig. 6. The reconstruction results of point-like phantom: (a) ground-truth; (b) delay-and-sum; (c) time reversal; (d) U-Net; (e) proposed Y-Net; (f) The profile along the white dotted line of (a), (b), (c), (d), (e).

$$\text{SSIM}(f, gt) = \frac{(2\mu_f \mu_{gt} + C_1)(2\sigma_{cov} + C_2)}{(\mu_f^2 + \mu_{gt}^2 + C_1)(\sigma_f^2 + \sigma_{gt}^2 + C_2)}, \quad (20)$$

where $\mu_f$, $\mu_{gt}$ and $\sigma_f$, $\sigma_{gt}$ are the local means, and standard deviations of $f$ and $gt$ respectively, and $\sigma_{cov}$ is cross-covariance for $f$, $gt$.

2). The Peak Signal-to-Noise Ratio (PSNR) is a conventional metric of the image quality in decibels (dB):

$$\text{PSNR}(f, gt) = 10\log_{10}(\frac{I_{max}^2}{MSE}), \quad (21)$$

where $I_{max}$ is the max value of $f$, $gt$ (in this work, $I_{max}=1$), $MSE$ can be calculated by Eq. (17).

3). The Signal-to-Noise Ratio (SNR) is defined as the ratio of peak signal intensity and standard deviation of the background intensities in decibels, which is only based on signal and noise:

$$\text{SNR}(f) = 10\log_{10}\left(\frac{peak(f)}{\sigma_b}\right)^2, \quad (22)$$

where $peak(f)$ is the peak intensity of $f$, $\sigma_b$ is the standard deviation of background. We also compare two variant Y-Net with our approach, which removes the connection of raw data (Encoder I) and the connection of the beamformed image (Encoder II) with the Decoder respectively. Meanwhile, the post-processing method based U-Net that only input an image after beamforming is also demonstrated for evaluation.

TABLE I
QUANTITATIVE EVALUATION OF DIFFERENT METHODS FOR TEST SETS

| Algorithms | SSIM | PSNR | SNR |
|---|---|---|---|
| delay-and-sum (DAS) | 0.2032 | 17.3626 | 1.7493 |
| time reversal (TR) | 0.5587 | 17.8482 | 2.2350 |
| Y-Net (concatenate Encoder II) | 0.8988 | 25.2708 | 9.6577 |
| Y-Net (concatenate Encoder I) | 0.8622 | 23.9152 | 8.105 |
| U-Net | 0.9002 | 25.0032 | 9.3233 |
| proposed Y-Net | **0.9119** | **25.5434** | **9.9291** |

The performance comparison of the test set is shown in TABLE I. Firstly, the deep learning based methods show obviously advantageous than conventional algorithms. Within the deep learning based approaches, the proposed network's performance is superior in comparison with the other networks.

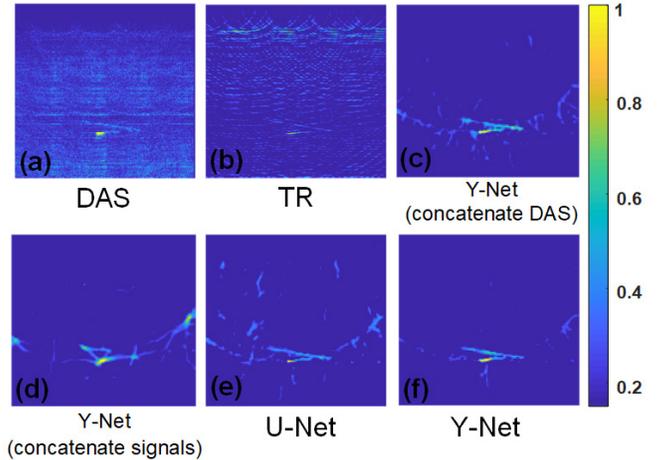

Fig.7. The *in-vitro* result of chicken breast phantom: (a) delay-and-sum; (b) time reversal; (c) Y-Net only concatenate Encoder II into the Decoder; (d) Y-Net only concatenate Encoder I into the Decoder; (e) U-Net; (f) Y-Net.

*B. Evaluation of experimental data*

The *in-vitro* results are shown in Fig. 7, which also compared DAS, TR, and two variant Y-Net and U-Net with Y-Net. DAS and TR methods show poor quality due to the laser power limit and severe artifacts (Fig. 7(a)-(b)), even though we still can distinguish the phantoms in the tissue. Deep learning based methods show higher SNR in Fig. 7(c)-(f). It shows that the Y-Net (concatenate signals) (Fig. 7 (d)) reconstructed an incorrect image, which completely lost the shape of phantom. The skipped connection between Encoder II and Decoder is the main reason, which provides a texture feature of the sample. The phantom's texture is different from vessel, and it causes the network to think of all signals as vessel-shape if lacking effective texture in Encoder II. U-Net removes most artifacts and retains some artifacts in extension direction, which embodies the associative ability. Y-Net shows a better result that can clearly distinguish the object (Fig. 7(f)).

The *in-vivo* imaging results comparison is shown in Fig. 8, where the ROI is limited by spot size. DAS and TR methods reconstructed images show many artifacts in tissue (Fig.8 (a)-(b)), but the major vessel can be recognized. Deep learning based methods have unsatisfactory results on the shape of the blood vessels due to an excessive association, especially in Fig.8 (d). These models can eliminate most noise and artifacts. The bottom right corner may be a vessel in deeper tissue or an intense artifact. U-Net removes normal artifacts and connects two vessels based on the extend tendency of the vessel, which is caused by excessive association (Fig. 8(e)). However, Y-Net still showed good performance, with no excessive associations on the main blood vessel (Fig.8 (f)).

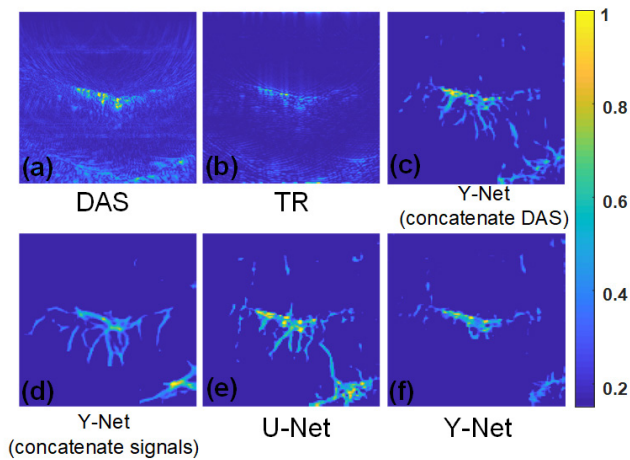

Fig.8. The *in-vivo* result of human palm: (a) delay-and-sum; (b) time reversal; (c) Y-Net only concatenate Encoder II into the Decoder; (d) Y-Net only concatenate Encoder I into the Decoder; (e) U-Net; (f) Y-Net.

The computation time for each method has been listed in TABLE II, which sufficiently satisfies the requirement of real-time imaging for most applications. Be specific, DAS and TR are implemented by MATLAB, and deep learning methods are implemented by Python.

TABLE II
THE COMPUTATION TIMES FOR EACH METHOD

| Algorithms | Time (Second) |
|---|---|
| delay-and-sum (DAS) | 0.25 |
| time reversal (TR) | 2 |
| Y-Net (concatenate Encoder II) | 0.0309 |
| Y-Net (concatenate Encoder I) | 0.0299 |
| U-Net | 0.0189 |
| proposed Y-Net | 0.0326 |

## VI. DISCUSSION

The artifacts are essential to limited-view photoacoustic tomography. An effective strategy to reduce artifacts is a challenge in image reconstruction. The model-based methods incorporate the physical model into the reconstruction process with a regularization, such as total variation (TV), and it also shows powerful performance. However, the non-iterative methods with Deep Learning are promising for applications where low latency is more important than a better quality reconstruction, such as real-time imaging for cancer screening and guided surgery.

In the comparative experiment, we chose U-Net as post-processing reconstruction scheme, which has been proven to work well on medical image reconstruction. In the experiment, the reconstruction results are deviated due to the gap between simulation data and measurement data, but our method still shows better performance compared to other methods.

## VII. CONCLUSION

In this paper, a new CNN architecture, named Y-Net, is proposed, which consists of two intersecting encoder paths. The Y-Net takes two types of inputs that represent the texture structure of the conventional algorithms and the high-dimensional features contained in the original raw signals respectively. We use the k-Wave PA simulation tool to generate a large amount of training data to train the network, and evaluate our approach on the test set. In the experiment, we demonstrate the feasibility and robustness of our proposed method by comparing with other models and conventional methods. We also validated our method in *in-vitro* and *in-vivo* experiments, showing superior performance compared with existing methods. Y-Net is still affected by the artifacts of beamforming, which may be improved by using a better beamforming algorithm. In the future work, we will further generalize Y-Net to three dimensions for real-time 3D PA imaging.

ACKNOWLEDGMENT

This research was funded by Natural Science Foundation of Shanghai (18ZR1425000), and National Natural Science Foundation of China (61805139).